\documentclass[a4paper,10pt]{preprint}
\usepackage{times}
\usepackage{mhequ}
\usepackage{epsfig}
\usepackage{amssymb}
\newtheorem{theorem}{Theorem}[section]

\newtheorem{proposition}[theorem]{Proposition}

\newtheorem{remark}[theorem]{Remark}
\newtheorem{conjecture}[theorem]{Conjecture}

\newcommand{\proba}{{\bf P}}
\newcommand{\proof}{\hfill\break\noindent{\bf Proof}. }

\newcommand{\tri}{ \mathcal{T}}
\newcommand{\noeuds}{\mathcal{N}}
\newcommand{\lienst}{\mathcal{L}(T)}
\def\eg{{\it e.g.}}

\def\dlap#1{{\setbox0=\hbox{#1}\dimen0=\ht0\divide\dimen0by
1\raise-\dimen0\box0}}
\def\ulap#1{{\setbox0=\hbox{#1}\dimen0
=\ht0\divide\dimen0
by 1\raise\dimen0\box0}}
\def\nlap#1{{\setbox0=\hbox{#1}\dimen0
=\ht0\divide\dimen0
by 2\raise-\dimen0\box0}}
\def\figureeewithtex#1#2#3#4#5#6#7#8#9{
\def\myunit{1cm}
\setlength{\unitlength}{\myunit}
\def\mywidth{#1}
\def\myheight{#2}
\def\myxorig{7.5}
\def\myyorig{5}
\def\picoffset{#3}
\def\mytxoff{#4}
\def\mytyoff{#5}

\begin{figure}\label{fig:#9}
\C{\psfig{figure=#6}}%
\leavevmode
\input{#7}
\end{picture}%
\null\vskip\picoffset cm
\caption{#8}
\end{figure}
}%

\newbox\MHbox
\def\figurewithtex#1#2#3#4#5#6#7#8#9{
\def\myunit{1cm}
\setlength{\unitlength}{\myunit}
\def\mywidth{#1}
\def\myheight{#2}
\def\myxorig{7.5}
\def\myyorig{#3}
\def\picoffset{#3}
\def\mytxoff{#4}
\def\mytyoff{#5}
\begin{figure}[ht]
\leavevmode\raise0\unitlength\vbox to\myheight\unitlength{%
\vfill\hbox to\mywidth\unitlength{%
\hfill%
\newdimen\figcenter
\figcenter=\hsize\relax
\divide\figcenter by 2%
\begin{picture}(-15,0)(\myxorig,\myyorig)%
\put(0,0){\makebox(21,10)[lb]{\epsfig{figure=#6}}}%
\put(0,\mytyoff){\makebox{\input{#7}}}%
\end{picture}%
\hfill}%
\vfill}%
\vspace{\picoffset\unitlength}

\caption{#8}\label{#9}
\end{figure}
}%
\def\newfigurewithtex#1#2#3#4#5#6#7#8#9{
\def\myunit{1cm}
\setlength{\unitlength}{\myunit}
\def\mywidth{#1}
\def\myheight{#2}
\def\myxorig{7.5}
\def\myyorig{#3}
\def\mytxoff{#4}
\def\mytyoff{#5}
\begin{figure}[ht]
\leavevmode\raise0\unitlength\vbox to\myheight\unitlength{%
\vfill\hbox to\mywidth\unitlength{%
\hfill%
\newdimen\figcenter
\figcenter=\hsize\relax
\divide\figcenter by 2%
\begin{picture}(-15,0)(\myxorig,\myyorig)%
\put(0,0){\makebox(21,10)[lb]{\epsfig{figure=#6}}}%
\put(0,\mytyoff){\makebox{\input{#7}}}%
\end{picture}%
\hfill}%
\vfill}%
\vspace{\picoffset\unitlength }
\caption{#8}\label{#9}
\end{figure}
}%

\let\epsilon=\varepsilon

\def\ie{{\it i.e.}}
\begin{document}
\title{Dynamics of Triangulations}

\author{P. Collet${}^1$, J.-P. Eckmann${}^{2,3}$}
\institute{${}^1$Centre de Physique Th\'eorique,\\ CNRS UMR 7644,
Ecole Polytechnique,
F-91128 Palaiseau Cedex (France)\\
${}^2$D\'epartement de Physique Th\'eorique, Universit\'e de
Gen\`eve,\\${}^3$Section de Math\'ematiques, Universit\'e de
Gen\`eve}

\maketitle

\begin{abstract}
We study a few problems related to Markov processes of flipping
triangulations of the sphere. We show that these processes are ergodic
and mixing, but find a natural example which does not satisfy detailed
balance. In this example, the expected distribution of the degrees of
the nodes seems to follow the power law $d^{-4}$.
\end{abstract}
\thispagestyle{empty}

\section{Introduction}\label{Introduction}
 We consider a Markov chain on triangulations of
the sphere (or other surface). Let $\tri$ denote the set of
triangulations, by this we mean the set of all combinatorially
distinct rooted simplicial 3-polytopes.

Tutte \cite{Tutte1962} showed that their number is asymptotically
\begin{equation}\label{Tutte1962}
Z_n\,=\, \frac{3}{16\sqrt{6\pi n^5}}\left( \frac {256}{27}\right)
^{n-2}~,
\end{equation}
as the number $n$ of vertices goes to $\infty$.
Of course, Euler's theorem holds for such triangulations, and this
means that when there are $n$ nodes, there are also $3n-6$ links and
$2n-4$ triangles.

For an element $T\in \tri$, we denote by $\noeuds(T)$ the set
of nodes and by $\lienst$ the set of links.

For any link $\ell$ (connecting the nodes $A$ and $B$), we consider the
``complementary'' link $\ell'$, which is defined as follows:
if $(A,B,C)$ and $(A,B,D)$ are the two triangles sharing the link
$\ell$, then $\ell'$ is the link connecting $C$ and $D$.

We assume that for any $T\in \tri$, a probability $\proba_{T}$ is
given on $\lienst$, \ie, $\sum_\ell \proba_T(\ell)=1$.
We define a Markov chain on $\tri$ as
follows. We first choose a link $\ell\in \lienst$ at random (with
probability $\proba_T(\ell)$). 

$\bullet$ If the link $\ell'$ belongs to $\lienst$, we do not change $T$
 and proceed with the next independent choice of a link.

$\bullet$ If $\ell'$ does not belong to $\lienst$, we erase $\ell$ and
replace it by $\ell'$. We obtain in this way a new triangulation $T'$
and we proceed with the next  independent choice of a link. This
replacement of $\ell$ by $\ell'$ 
is commonly called a {\em flip}
see \cite{Negami1999},  or a Gross-Varsted move \cite{Malyshev1999}.
See Fig.~\ref{f:tree0}.
\def\picoffset{-1}
\newfigurewithtex{2}{5}{5}{0}{0}{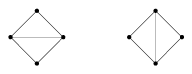}{christmas0.inp}
{A flip: the link (A-B) is exchanged with (C-D).}{f:tree0}

We will denote by $\proba(T'|T)$ the transition probability of this
Markov chain.
\section{Properties of the Markov chain}

We now fix $n$ and let $\tri_n$ denote those triangulations with $n$ nodes.

\begin{proposition}\label{ergo}
Assume that
$\inf_{T\in\tri_n}\inf_{\ell\in\lienst}\proba_{T}(\ell)>0$. Then the Markov
chain defined in  Sec.~\ref{Introduction} is irreducible and aperiodic.
\end{proposition}
\proof It is well known (see \cite{Negami1999}) that by flipping links as
described above one can connect any two triangulations
of $\tri_n$ (one shows that any $T$ can be flipped a finite number of
times to reach a
``Christmas tree'' configuration). 
\def\picoffset{0.5}
\newfigurewithtex{2}{8}{5}{0}{0}{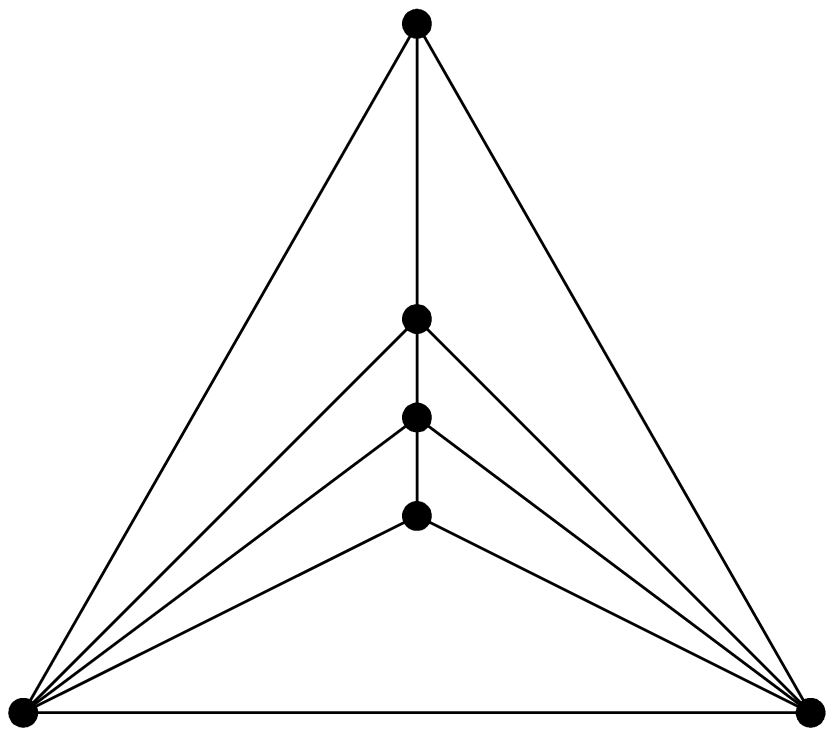}{christmas1.inp}
{The ``christmas tree'' with $n$  nodes, the ``branches'' between 5
  and $n$ not being shown. Any triangulation can be brought to this form
by a sequence of flips.}{f:tree}

Since by
our hypothesis any such (finite) succession of moves has a non-zero
probability this shows the irreducibility of the chain. To prove
aperiodicity, we have to prove that for any  high enough iterate of the
transition matrix, all the diagonal entries are positive. By the previously
mentioned result, it is enough to show that we can construct cycles of
length two and three for the  ``Christmas tree''. Cycles of length two
are easily obtained by flipping a link back and forth. For cycles of
length three, we consider the sub ``Christmas tree'' of size six at the
base of the complete ``Christmas tree'', see Fig.~\ref{f:tree}.
We enumerate the nodes as in the figure, assuming $n\ge 7$.
In
particular node 3 has degree 3, nodes 4 to $n$ have degree 4 and nodes 1
and 2 degree $n-2$.
\def\picoffset{-0.5}
\newfigurewithtex{-1}{4}{4}{0}{0}{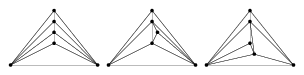}{christmas2.inp}
{The three stages in the cycle of 3 flips regenerating the christmas
  tree with $n$ nodes.}{f:tree2}

The cycle of
length 3 is obtained by performing the following flips
$$
\begin{array}{ccc}
(1-4)&\to&(3-5)\cr
(2-3)&\to&(1-4)\cr
(4-5)&\to&(2-3)\cr
\end{array}
$$
after which we get again the ``Christmas tree'' with nodes 3 and 4 exchanged.
\hfill$\square$

\begin{remark}\rm
Of course, the condition of Prop.~\ref{ergo} is not necessary, but we
do not know any simple other criterion in terms of the $\proba_T$, but
one can think for example of conditions involving two successive flips.
\end{remark}

From this result we conclude that there is only one invariant
probability measure, and with this measure the chain is ergodic and mixing.

\section{Two Examples}

The easiest example is that where one chooses a link uniformly at
random. Then one gets the uniform
distribution on $\tri$, and, using this simple fact, many properties
of this process can be deduced, see, {\it e.g.}, \cite{Godreche1992}.

Here, we consider another example, which was suggested to us by
Magnasco \cite{Magnasco1992,Magnasco1994}. This process consists in first choosing a node
uniformly and then to choose uniformly a link from this node. Let $n$ be
the number of nodes.  An easy computation, shown below, leads to
\begin{equation}\label{probalien}
\proba_{T}(\ell)=\frac{1}{n}\left(\frac{1}{d_{1}(\ell|T)}
+\frac{1}{d_{2}(\ell|T)}\right)~,
\end{equation}
where $d_{1}(\ell|T)$ and $d_{2}(\ell|T)$ are the degrees of the nodes  
at the ends
of link $\ell$ in the triangulation $T$.

\proof If $\ell$ is a link, we denote by $\partial \ell$ the two nodes it
connects. If $\ell$ is a link and $i$ is a node, we say that $\ell\sim
i$ if $i\in\partial\ell$.  If $i$ is a node, we denote by $d_{i}(T)$ its
degree in the triangulation $T$.
We have from Bayes' formula
$$
\proba_{T}(\ell)=
\sum_{i\in\noeuds}\proba_{T}(\ell\,|\,i)\proba(i)\;.
$$
Moreover, $\proba(i)=1/n$ for any $i$, $\proba_{T}(\ell\,|\,i)=0$  if
$\ell\not\sim i$ and otherwise
$$
\proba_{T}(\ell\,|\,i)=\frac{1}{d_{i}(T)}\;.
$$
Therefore
$$
\proba_{T}(\ell)=\frac{1}{n}
\sum_{i\in\partial\ell}\frac{1}{d_{i}(T)}
$$
which is formula (\ref{probalien}).\hfill$\square$

It also follows directly from this expression 
that for any $T\in \mathcal{T}$,
\begin{equs}
\sum_{\ell\in T}\proba_{T}(\ell)&=\frac{1}{n}\sum_{\ell\in T}
\sum_{i\in\partial\ell}\frac{1}{d_{i}(T)}\\
&=\frac{1}{n}
\sum_{i}\frac{1}{d_{i}(T)}\sum_{\ell\in \mathcal{L}(T)\,,\, \ell\sim i}1=
\frac{1}{n}
\sum_{i}\frac{1}{d_{i}(T)}d_{i}(T)=1\;.
\end{equs}
In this computation we have not used the fact that $T$ is a
triangulation. Therefore this relation holds for any graph.

For the second model, we have
\begin{theorem}
The Markov chain $\proba(\,\cdot\,|\,\cdot\,)$ is not reversible (when
$n\ge7$).
\end{theorem}

\noindent{\bf Remark.} {\rm We have not checked what happens for smaller $n$.}

In other words, one cannot easily guess the invariant measure from the
transition probabilities.
\proof Assume the chain is reversible with respect to some probability
$\proba$ on $\tri$, namely  for any $T$ and $T'$ in $\tri$ we have
\begin{equation}\label{rever}
\proba(T'|T)\proba(T)=\proba(T|T')\proba(T')\;.
\end{equation}
If $T_{1},\ldots,T_{k},T_{k+1}=T_{1}$ is any cycle of admissible flips, we
must have
$$
\prod_{j=1}^{k}\frac{\proba(T_{j}|T_{j+1})}{\proba(T_{j+1}|T_{j})}=1\;.
$$
We are going to show that there is a cycle of length 4 for the
christmas graph for which this is not true,
see Fig.~\ref{f:tree3}. 
\def\picoffset{1.2}
\newfigurewithtex{-8}{7.5}{1}{0}{0}{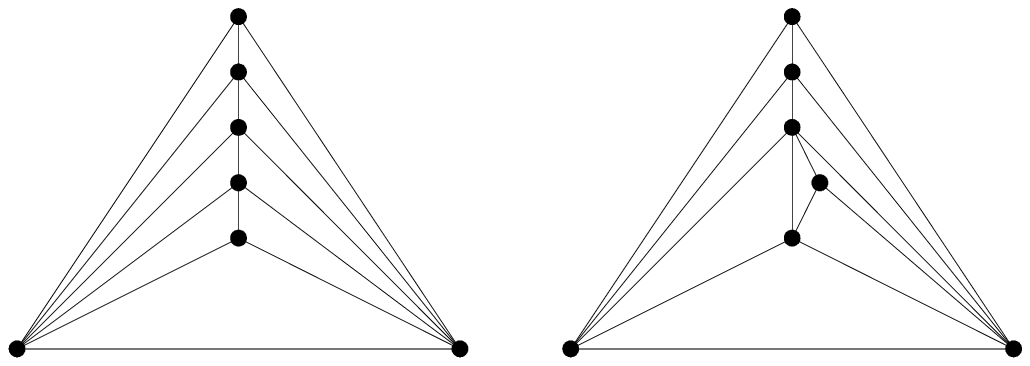}{christmas3.inp}
{The four stages of the cycle of 4 flips which regenerate the christmas
  tree with $\ge7$ nodes, but which show the {\em absence} of detailed
  balance. (Top left $\to$ top right $\to$ bottom left $\to$ bottom
  right $\to$ top left.)}{f:tree3}

Consider the following cycle for the ``Christmas tree'' with the same notations as
before 
$$
\begin{array}{ccc}
(1-4)&\to&(3-5)\cr
(2-5)&\to&(4-6)\cr
(3-4)&\to&(2-5)\cr
(5-6)&\to&(1-4)\cr
\end{array}
$$
An easy computation leads to
$$
\prod_{j=1}^{4}\frac{\proba(T_{j}|T_{j+1})}{\proba(T_{j+1}|T_{j})}=\frac{10}{9}\;,
$$
if the number of nodes is larger than 6.\hfill$\square$

\section{Numerical simulation}\label{simul}
\begin{figure}
\psfig{file=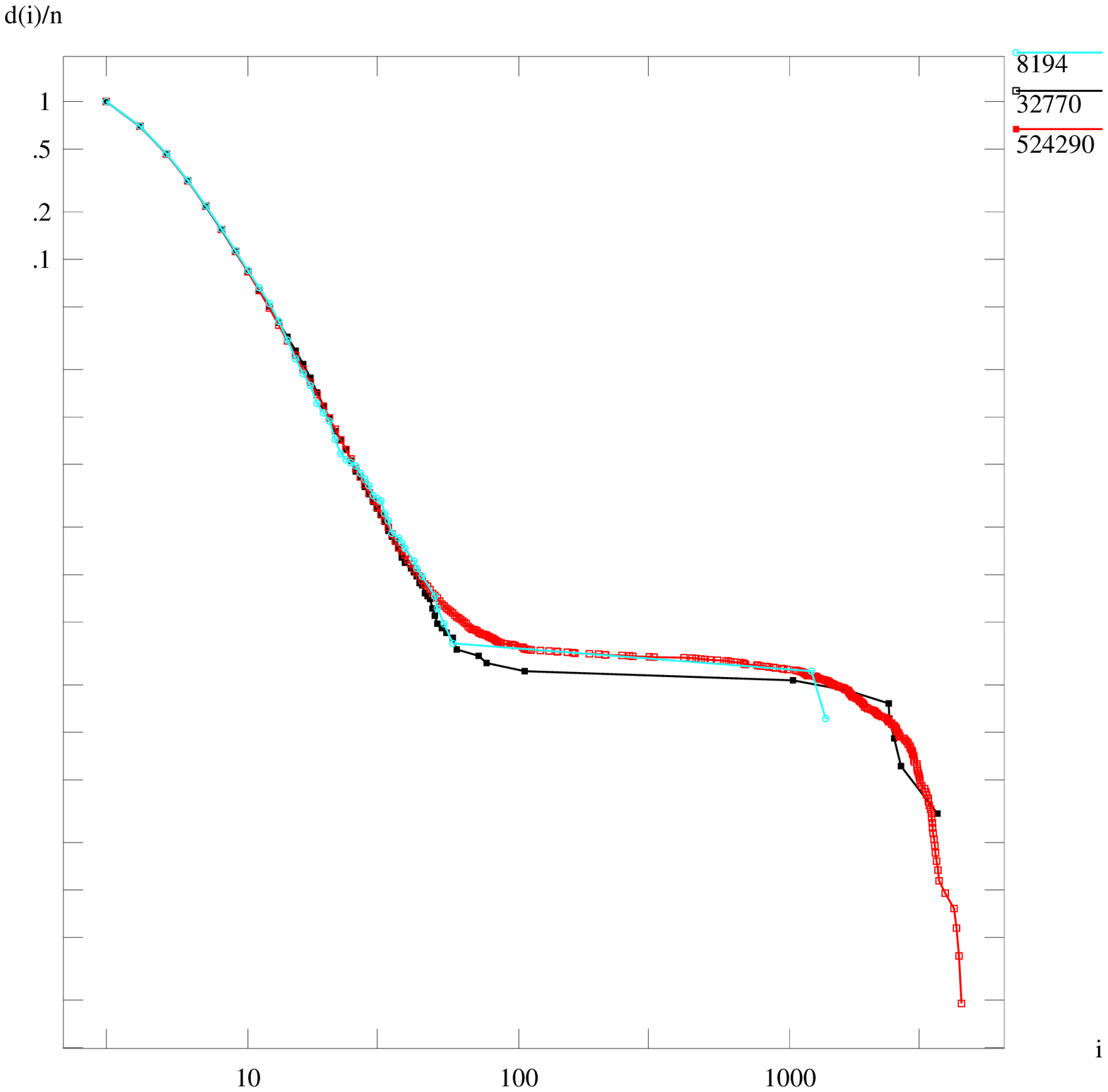,width=10cm}
\caption{A log-log plot for triangulations of size $n=8194$, 32770, and
  524290, after about $10^{10}$ flips. The data are the cumulated sum $d(i)$
  of number of nodes with degree $\ge i$. The straight part is well
  fitted with a law of $d(i)\sim c d^{-3}$, so that the degree
  distribution seems to be $\sim d^{-4}$. The outliers are produced by
  the lacunarity of the data when the expected number of nodes of a
  given 
  degree starts to be less than 1.  }\label{fig1}
\end{figure}

We have performed extensive simulations on the model described
above. In this section we summarize the numerical findings, but the
reader should note that we have no theoretical explanation for the
results. The main insight is that the model with the {\em uniform}
measure \cite{Godreche1992} leads to an exponential degree distribution,
while the model of \cite{Magnasco1994} leads to a power law distribution in
a sense which we make clear now, see Fig.\ref{fig1}. 

We formulate the
results as
\begin{conjecture}\label{conj} There is a probability measure $p$ on the
integers larger than 2 such that the average number of nodes of degree
$d$ divided by $n$ converges  when $n$ tends to infinity to
$p(d)$. Moreover $p$ has polynomial decay in the sense that $d^{-4}p(d)$
converges to a nonzero finite limit when $d$ tends to infinity.
\end{conjecture}
\begin{remark}\rm
It should be noted that several deviations from a pure power law are
present in these experiments and will not go away with large
$n$. First of all, nodes of degree 3 are less frequent than would be
suggested by a power law. We attribute this to the impossibility of
doing a flip if a node of degree 3 is chosen: All its edges are
unflippable.
Second, if there are $n$ nodes, assuming an approximate power law of 
$N(d)=c\cdot d^{-4}$ we find  $c\approx 50n$ (from $c\sum_{d=3}^\infty=n$).
Thus there should be no nodes for which $N(d)<1$, that is
$50nd^{-4}<1$ or $d>(50n)^{1/4}$. However, the experiments clearly
show the presence of 
``outliers'' of much larger degree. Closer analysis (with, \eg,
logarithmic binning) reveals that these
outliers are spaced at equilibrium in a way to {\em continue} the
measured power law.
\end{remark}

We have done extensive checks for the correlations of degrees of
neighboring nodes in our simulations. Such correlations have been
theoretically explained in \cite{Godreche1992} for the case of the uniform
choosing rule. These correlations are difficult to measure, but no
decisive deviation from independence was found, except for some obvious
topological rules.

There is a feeling in the community of specialists, be they interested
in random triangulations, or in 2-d gravity (the dynamic dual of our
problem) that the ``typical'' triangulation should be ``flat'' (which
means that each node should (wants to?) have 6 links). 
To measure the effect of the tails of distribution of degrees, we
use
combinatorial differential geometry, as advocated by Robin Forman \cite{Forman2003},
who introduces a notion of ``combinatorial Ricci curvature'' which, in our
case of triangulations reduces to $\sum_i d_i^2 -5d_i$. Extensive
simulations show that this quantity seems to grow more or less
monotonically as the process reaches the equilibrium state. Note that
since $\sum_i d_i$ does not depend on the triangulations, we are just
measuring the sum of the squares of the degrees. 

Another observation, which holds with very high accuracy is that once
a node has been chosen, at equilibrium, exactly 50\% of all attempted
flips are {\em not} possible, because the ``other'' link is already
present. This means that a tetrahedron is placed on top of a
triangle. Note that the study of such ``vertex-insertions'' is already
present in Tutte's work \cite{Tutte1962}.
\bibliographystyle{JPE}
\markboth{\sc \refname}{\sc \refname}
\bibliography{refs}

\end{document}